\documentclass[11pt,technote,a4paper,onecolumn,twoside]{IEEEtran} 

\usepackage{times}

\usepackage{moreverb}

\usepackage{amsmath, amssymb}

\usepackage{multicol}

\usepackage{epsfig}
\usepackage{cite}
\usepackage{url}

\usepackage{theorem}
\theorembodyfont{\normalfont}

\makeatletter 
\allowdisplaybreaks[4]
\def\@oddfoot{\textbf\footnotesize \hfil\thepage}
\def\@evenfoot{\textbf\footnotesize \thepage \hfil}
\def\QEDclosed{\mbox{\rule[0pt]{1.3ex}{1.3ex}}} 

\def\QED{\QEDclosed} 

\makeatother

\title{Effect of payload size on mean response time
when message segmentations occur
using $\rm{M}^{\rm X}/\rm{G}/1$ queueing model
}

\author{Takashi Ikegawa
\thanks{T.~Ikegawa is with Waseda Research Institute for Science and Engineering,
Waseda University and 
Graduate School of Mathematical Sciences, the University of Tokyo, Japan
(e-mail: ikegawa@aoni.waseda.jp or tikegawa@ms.u-tokyo.ac.jp).}
}

\begin{document}

\maketitle

\begin{abstract}

This paper proposes the $\rm{M}^{\rm X}/\rm{G}/1$ queueing model
to represent arrivals of segmented packets when message segmentations occur.
This queueing model enables us to
derive the closed form of mean response time,
given payload size, message size distribution and message arrival rate.
From a numerical result,
we show that the mean response time is more convex in payload sizes
if message arrival rate is larger
in a scenario where Web objects are delivered over a physical link.

\end{abstract}


\section{Introduction}

The size of packets, i.e., data units transferred over physical links,
affects several quality of service (QoS) parameters
for users of packet-based transfer networks such as the Internet. 
For example, the response time between hosts is
highly dependent on packet size
because it contains link-level transmission time,
which is simply given by the packet size divided by capacity of the physical link \cite[p.~9]{KOB08_book}. 
In addition,
the packet-loss rate due to bit errors depends on the packet size
because it is approximately proportional to the packet size \cite[p.~132]{SCH87}. 

The packet size is limited for various reasons \cite[pp.~406--409]{TAN96}.
They include a) the data-link structure (e.g., the width of a transmission slot), 
b) compliance with standard protocol specifications,
and c) satisfaction of the QoS parameters by applications such that the round-trip time of interactive applications is less than the time that a user is willing to wait \cite{MON00, DAW01}.

Messages, i.e., data units generated by applications,
are frequently larger than the maximum permitted packet size. 
To convey such messages over the network, some communication protocols,
such as transmission control protocol (TCP), Internet protocol (IP) for the Internet \cite{STE94a}, IEEE 802.11 media access control (MAC) protocol \cite{CRO97} for wireless local-area networks (LANs) and radio link control (RLC) protocol \cite{RLC} for mobile wide-area networks, specify a message segmentation/reassembly function. 
The message segmentation function enables a sender to divide a single message larger than the payload size $\ell_d$ into multiple packets. 
Furthermore, the sender adds an appropriate header, i.e., overhead, to a packet.

There have been several studies on optimization of packet size (or payload size)
to satisfy the user's QoS such as \cite{LET98, MOD99, JEL08}.
However, the purpose of these studies was
to solve the tradeoff issue between the desire to reduce the header overhead by making packet large,
and the need to reduce packet-loss rate due to bit errors in noisy links
by using small packet size.

Another tradeoff exists when message segmentations occur.
Consider the message whose variance is large enough in size,
such as Web files (objects).
While too large payload size is employed,
message segmentations hardly occur.
In this case, the waiting times of packets for transmission using a physical link
is large because the packet size distribution can be identified with the message size distribution.
On the other hand,
when payload size is small,
waiting times of packets may be small
due to decreasing the variance of packet sizes
because the constant packets in size, which is payload size,
are dominant in the all created packets.
When payload size is too small,
the number of segmented packets per message increases significantly,
resulting in very large waiting time due to the burstiness of packet arrivals.

In previous work,
this tradeoff issue has not been discussed.
The purpose of this paper is to 
discuss the effect of payload size on mean response time
when message segmentations occur
using an $\rm{M}^{\rm X}/\rm{G}/1$ queueing model \cite[Chapter~4]{AKI99},
which can capture the behavior of the burstiness of segmented packet arrivals.

The rest of the paper is organized as follows.
In the next section, we describe the communication network model
underlying our study.
Section~\ref{sec: packet sequence} analyzes the segmented packet sequence.
Section~\ref{sec: R} derives the closed form of mean response time
using an $\rm{M}^{\rm X}/\rm{G}/1$ queueing model.
Section~\ref{sec: results} investigates the effect of message segmentation
on the mean response time for actual message size distributions.
Finally, Section~\ref{sec: conclusion} summarizes this paper and mentions future work. 

\section{Communication network model}

\begin{figure}
\centering
\epsfig{file=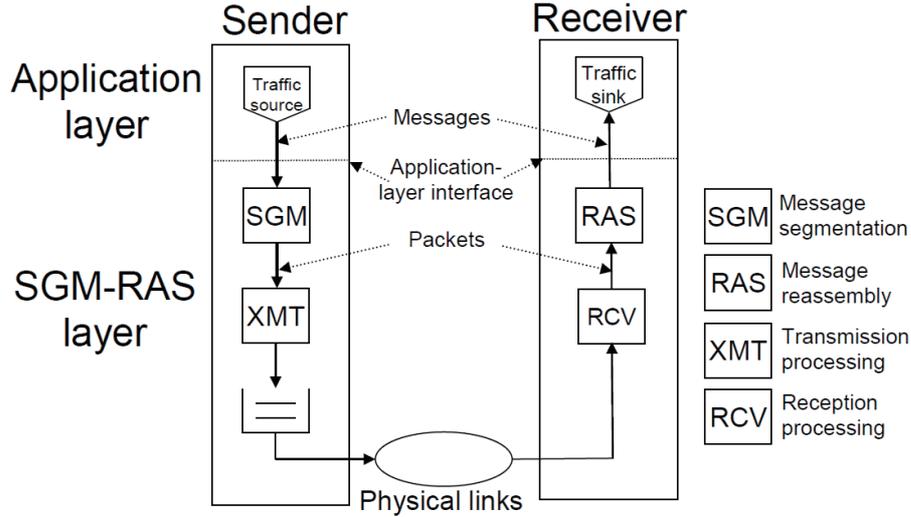, width=12cm} \\
\caption{Communication network model.}
\label{fig: network}
\end{figure}

In this section,
we first explain the two-layered communication network model
under consideration.
Next,
the model of data units introduced in this paper
at the respective layer is described.
In final,
we explain some assumptions for analytical tractability.

\subsection{Layer model}

To investigate the effect of message segmentation on performance,
we consider a communication network
of which conceptual representation is shown in Fig.~\ref{fig: network}.
Each station (a sender and a receiver) has two layers:
application and segmentation-reassembly (SGM-RAS).

The application layer contains a traffic source and sink.
The traffic source generates the data units,
which will be referred to as messages.
On the other hand,
the traffic sink terminates the corresponding data units.

The SGM-RAS layer implements message SGM-RAS function.
In addition,
it has a function to transfer data units,
which will be referred to as packets,
over physical links at a sender.

\subsection{Data-unit model}

We define data units exchanged between peer entities at the respective layer: messages and packets.

\begin{description}

\item{\bf Message:}
a data unit generated by a traffic source
with a given size distribution,

\item{\bf Packet:}
a data unit created from a message through segmentation function
by adding a header and/or trailer, i.e., control information,
to the (divided) message.
The message segmentation function implemented in the sender's SGM-RAS layer
enables a single message to be divided
into several packets
if the message size is larger than the
payload size $\ell_d (>0)$.
The receiver's SGM-RAS layer performs a message reassembly function,
thus reassembling the segmented generated packets before delivering them
to the application layer.

\end{description}

\subsection{Assumptions}

For analytical tractability,
we make the following assumptions.

\begin{description}

\item[\texttt{A1}:]
message sizes are mutually independent and 
identically distributed 
according to a common message-size distribution $F^{(m)}(\cdot)$.
The distribution $F^{(m)}(\cdot)$ has a finite mean value $\ell^{(m)}$,
which is referred to as the mean message size.

\item[\texttt{A2}:]
the finite variance $\{\sigma^{(m)}\}^2$ of the message-size distribution exists.

\end{description}

\section{Segmented packet sequence model}
\label{sec: packet sequence}

\begin{figure}
\centering{
\epsfig{file=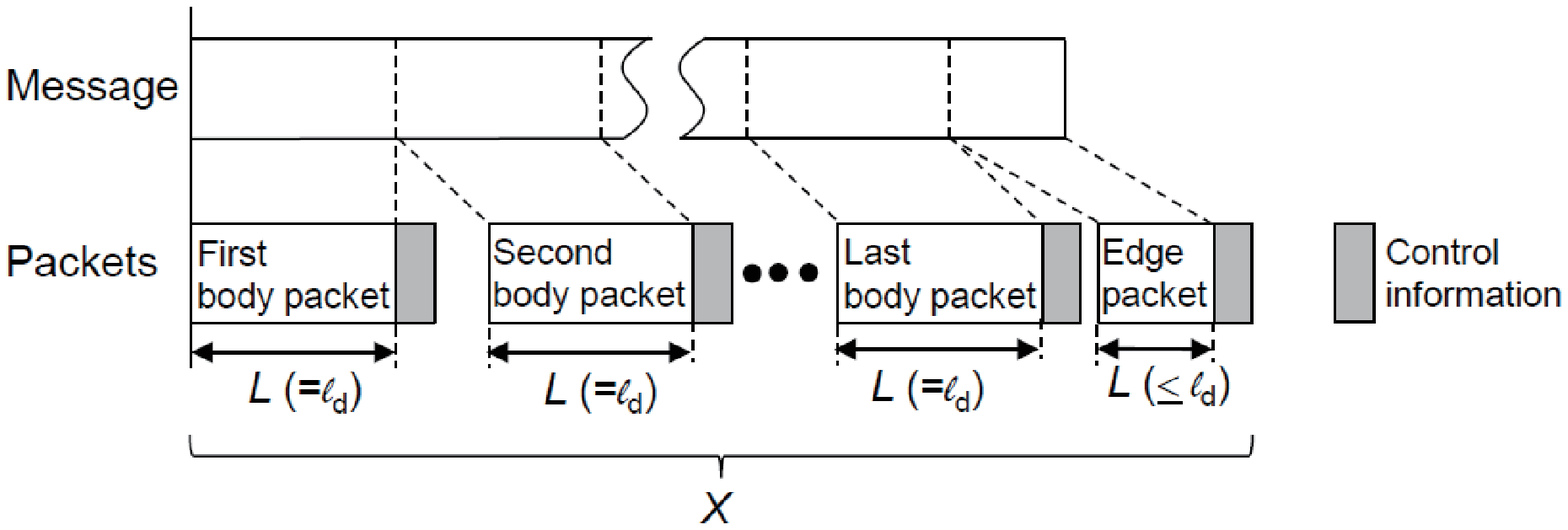, width=12cm} \\
\caption{Creation of packets through message segmentation.}
\label{fig: message segmentation}
}
\end{figure}

The creation of packets from a message through message segmentation is shown in Fig.~\ref{fig: message segmentation}.
If a message is larger than payload size $\ell_d$, the message is divided into multiple packets.
Two kinds of packets exist:
\begin{description}
\item{\bf body packet:}
a segmented packet appearing between the head and the penultimate packets in the original message, whose packet size is always equal to $\ell_d$, and
\item{\bf edge packet:}
the final segmented packet if a message is segmented, or the message itself if it is not segmented, whose packet size is variable but does not exceed $\ell_d$.
\end{description}

Let a random variable $L$ be a packet size excluding control information (header).
Letting $F^{(p)}(\cdot) \stackrel{\triangle}{=} \Pr. (L \le x)$ be the stationary distribution of packet sizes,
we have
\begin{align}
\label{eq: F^p}
F^{(p)}(x)
&= \left(1 - \pi^{(\textrm{E})}\right) \textbf{1}(x - \ell_d)
+ \pi^{(\textrm{E})} F^{(\textrm{E})}(x).
\end{align}
where $F^{(\textrm{E})}(\cdot)$ is the stationary edge-packet-size distribution and
$\pi_\textrm{E}$ is the edge packet occurrence probability.
The form of $F^{(\textrm{E})}(x)$ is given by
\begin{align}
\label{eq: F^E}
F^{(\textrm{E})}(x) 
&= \begin{cases}
0, & \text{if $x \le 0$}, \\
\displaystyle \sum_{n = 0}^{\infty}
\left\{F^{(m)}(x + n \, \ell_d) - F^{(m)}(n \, \ell_d)\right\},
& \text{if $0 < x \le \ell_d$}, \\
1, & \text{if $x > \ell_d$}.
\end{cases}
\end{align}
The form of $\pi^{(\textrm{E})}$ can be written as
\begin{align}
\pi^{(\textrm{E})} &= \cfrac{1}{\displaystyle\sum_{n=0}^\infty \, u_n}. 
\end{align}
Here, the term $u_n$ is defined as
\begin{align}
\label{eq: u_n}
u_n &\stackrel{\triangle}{=}
\int_{n \, \ell_d}^\infty dF^{(m)}(x) = 1 - F^{(m)} (n \,\ell_d),
& \text{for $n = 1, 2, \cdots$,}
\end{align}
with $u_0 = 1$, and we regard that $u_{n+1} / u_n=0$ when $u_n=0$.

Letting $\ell^{(p)}$ be the mean packet size,
from \eqref{eq: F^p} and assumption of \texttt{A1},
the form of $\ell_p$ is given by
\begin{align}
\ell^{(p)} &\stackrel{\triangle}{=} E\left[L\right] = \pi^{(\textrm{E})} \, \ell^{(m)}.
\label{eq: ell p}
\end{align}

Let $\{\sigma^{(p)}\}^2$ be the variance of packet sizes.
From \eqref{eq: F^p} and assumption of \texttt{A2},
the variance of packet sizes $\{\sigma^{(p)}\}^2$ is given by
\begin{align}
\label{eq: sigma_p^2}
\{\sigma^{(p)}\}^2 &\stackrel{\triangle}{=} E\left[L^2\right] - \left\{E\left[L\right]\right\}^2 \notag \\
&= \pi^{(\textrm{E})} \, 
\left(\left\{\ell^{(m)}\right\}^2 + \left\{\sigma^{(m)}\right\}^2\right) 
+ 2 \, \pi^{(\textrm{E})} \, \ell^{(m)} \, \ell_d \notag \\
&\quad
- 2 \, \pi^{(\textrm{E})} \, \ell_d 
\left(
\sum_{n = 0}^{\infty} \, v_n - \ell_d \, \sum_{n = 1}^{\infty} \, n \, u_n
\right) 
- \left\{\ell^{(p)}\right\}^2,
\end{align}
where
\begin{align}
\label{eq: v_n}
v_n &\stackrel{\triangle}{=} \int_{n \, \ell_d}^\infty \,x \,dF^{(m)}(x), 
& \text{for $n = 1, 2, \cdots$,}
\end{align}
with $v_0 = \ell^{(m)}$.

The detailed derivations of \eqref{eq: F^p} to \eqref{eq: v_n}
can be found in \cite{IKE12_PerEva}.

Let a random variable $X$ be the number of packets created from a corresponding message.
Note that the random variable $X$ is identified with a batch size,
which will be introduced in Section~\ref{sec: R}.

The forms of $E[X]$ and $E[X^2]$ are given by
\begin{align}
E[X] &= \sum_{n=0}^\infty \, n \, \Pr. \left(X = n\right)
= \sum_{n=1}^\infty \, n \, \left(u_{n-1} - u_n\right)
= \sum_{n=0}^\infty \, u_n \left(=\cfrac{1}{\pi^{(\textrm{E})}}\right),
\end{align}

\begin{align}
E[X^2] &= \sum_{n=0}^\infty \, n^2 \, \Pr. \left(X = n\right)
= \sum_{n=1}^\infty \, n^2 \, \left(u_{n-1} - u_n\right)
= \sum_{n=0}^\infty \, (2 \, n + 1) \, u_n.
\end{align}

\section{Mean response time}
\label{sec: R}

\begin{figure}
\centering{
\epsfig{file=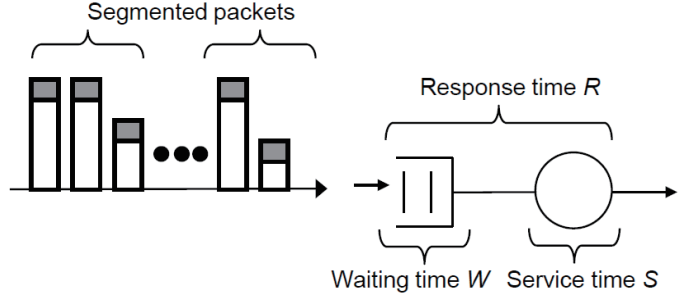, width=9cm} \\
\caption{Queueing system.}
\label{fig: queue}
}
\end{figure}

In this paper,
we employ the response time $R$ as the performance.
To derive the form of $E[R]$,
we represent a communication network as a queueing system as shown in Fig.~\ref{fig: queue}.
The response time $R$ is defined as the total time from a packet arrival at the queue until its service completion.
In the communication network,
a service time $S$ in a queueing system where the server is a physical link is equal to transmission time,
which is given by $\ell / \mu_c$ where a data unit is $\ell$ in size and $\mu_c$ is the capacity of the physical link.
The response time $R$ is interpreted as the waiting time $W$, i.e.,
time spent in the queue alone, plus the service time $S$.

For analytical tractability,
we introduce the following assumptions:
\begin{description}

\item[\texttt{B1}:]
packets created from a corresponding message, referred to as a batch,
arrive at the queueing system simultaneously at a time.

\item[\texttt{B2}:]
the batches arrive at the queueing system according to Poisson process with mean arrival rate $\lambda$.

\item[\texttt{B3}:]
the maximum number of packets that can be accommodated in the queueing system is infinite.

\item[\texttt{B4}:]
packets are served in FIFO scheduling dripline.

\item[\texttt{B5}:]
offered load $a = \lambda \, E[X] \, E[S]$ satisfies $a < 1$.

\item[\texttt{B6}:]
the size of SGM-RAS-layer's control information is constant
and equal to $\ell_h$.

\end{description}

From assumptions of \texttt{B1} -- \texttt{B4},
we can treat the queueing system as
an $\rm{M}^{\rm X}/\rm{G}/1$ queueing model in the Kendall notation,
where $X$ is the batch size,
which is defined as the number of packets simultaneously arriving at the queueing system.

From assumption of \texttt{B5},
the $\rm{M}^{\rm X}/\rm{G}/1$ becomes stable.

By solving the $\rm{M}^{\rm X}/\rm{G}/1$ queueing model,
we have
\begin{align}
\label{eq: R}
E[R] &= E[S] + E[R], \\
E[S] &= \cfrac{E[L] + \ell_h}{\mu_c}
= \cfrac{\ell^{(p)} + \ell_h}{\mu_c}, \\
\label{eq: S2}
E\left[S^2\right] &= 
	\cfrac{E\left[\left(L + \ell_h\right)^2\right]}{\mu_c^2} 
	= \cfrac{\left\{\sigma^{(p)}\right\}^2 + (\ell^{(p)} + \ell_h)^2}{\mu_c^2}, \\
\label{eq: W}
E[W] &=
\cfrac{
\lambda \, 
\left\{
E[X] \, E\left[S^2\right] 
+
\left(E\left[X^2\right] - E[X] \right) \, \left\{E[S]\right\}^2
\right\}
}
{2 \, (1 - a)} \notag \\
&\quad + \cfrac{\ell_d + \ell_h}{\mu_c} \left(E[X] - 1\right).
\end{align}

The derivations of \eqref{eq: R} -- \eqref{eq: W} can be found in Appendix.

\section{Numerical results and discussions}
\label{sec: results}

In this section,
we examine the effect of payload size on mean response time
by utilizing the results in Section~\ref{sec: R}.
We consider a scenario in which 
Web objects are transferred over the IEEE~802.11g,
i.e., physical link capacity $\mu_c = 54$~Mbps and
control information field size $\ell_h = 38$~bytes.

The sizes of the Web objects are assumed to follow a lognormal distribution given by
\begin{align}
\label{eq: F^{(m)}^l(x)}
F^{(m)}(x) &= 
\begin{cases}
\displaystyle \int_{y=0}^{y=x} \cfrac{1}{\sqrt{2 \pi} \sigma_w y}
e^{\tfrac{-\left(\log y - \mu_w\right)^2}
{2 \sigma_w^2}} dy, & x > 0, \\
0, & x \le 0.
\end{cases}
\end{align}

The distribution parameters $\mu_w$ and $\sigma_w$ are assumed to be $6.34$ and $2.07$, respectively,
on the basis of the measured mean message size $\ell^{(m)}=4827$ bytes and
the measured standard deviation $\sigma^{(m)}=41,008$~bytes \cite{MOL00}.
Note that this lognormal distribution can represent a long-tailed property.

\begin{figure}
\centering
\epsfig{file=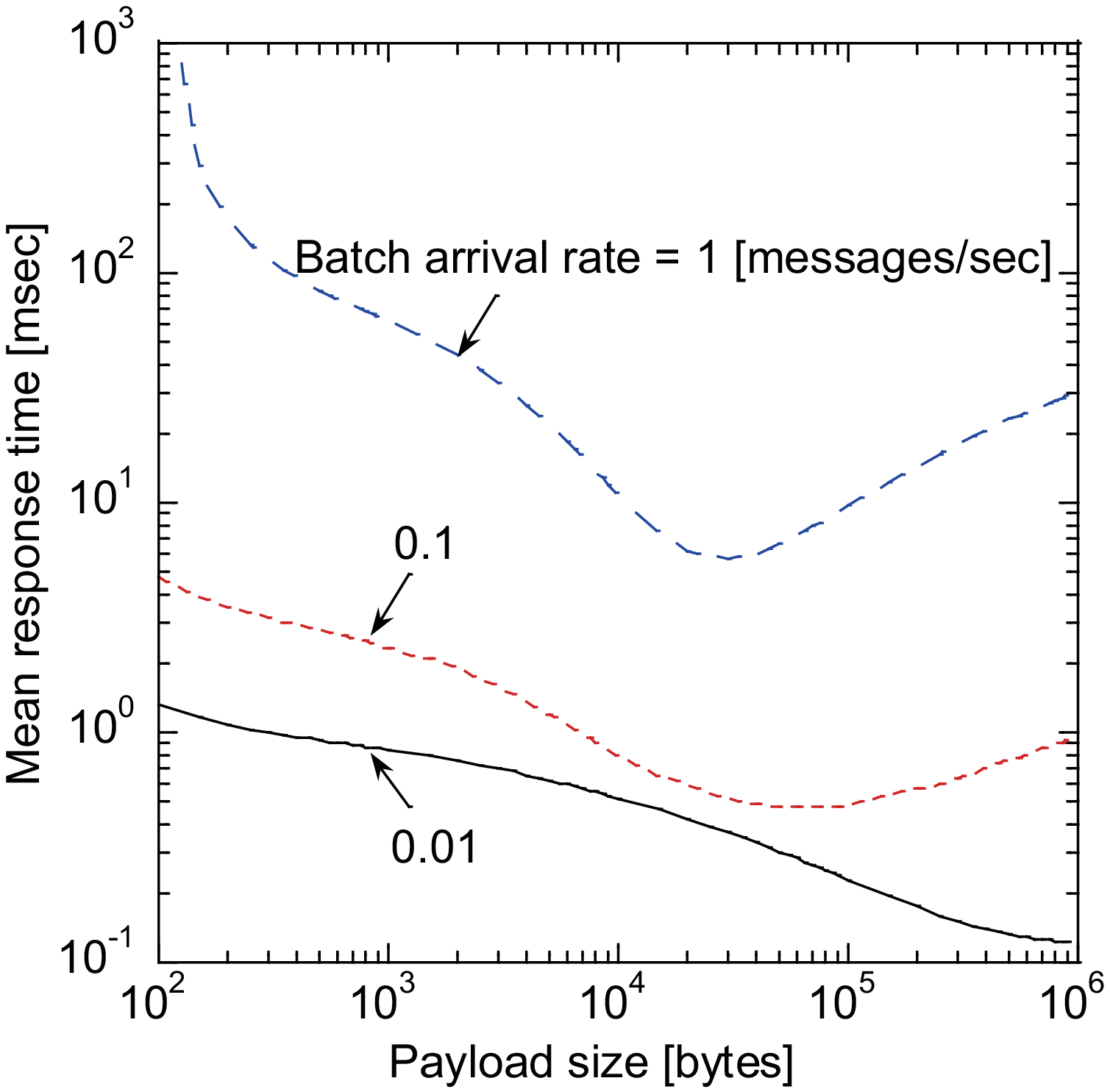, width=8cm} \\
\caption{Mean response time $R$ versus payload sizes $\ell_d$
for different batch arrival rates $\lambda$.}
\label{fig: log}
\end{figure}

Figure~\ref{fig: log} shows mean response time $R$ versus payload sizes $\ell_d$
for different batch arrival rates $\lambda$.
This figure demonstrates the mean response time $R$ is convex in payload sizes $\ell_d$.

The reason for this is as follows.
\begin{itemize}

\item {\bf When payload size $\ell_d$ is large enough}

From \eqref{eq: u_n}, $u_n$ for $n = 1, 2, \cdots$ is approximately zero,
resulting in $\pi^{(\textrm{E})} \approx 0$.
Hence, message segmentations hardly occur.

\ \ In addition, it yields $E[X] \approx 1$ and $E[X^2] \approx 1$.
Hence, since batch size is approximately one,
the queueing system can be approximated as an $\rm{M}/\rm{G}/1$ queueing model

The mean waiting time $E[R]$ for the $\rm{M}/\rm{G}/1$ queueing model is given by
\begin{align}
E[W] \approx 
\cfrac{\lambda \, E[S]^2}
{2 \, (1 - \lambda \, E[S])} 
= \cfrac{
\lambda \, \{E[S]\}^2
\left\{
\left(
\cfrac{
\sqrt{E[S^2] - \left\{E[S]\right\}^2}}{E[S]}
\right)^2 + 1
\right\}
}
{2 - \lambda \, E[S]}.
\end{align}

\ \ From the above equation,
we find that mean response time $R$ depends on the coefficient of variation of $S$, $c_S$,
that is, $\sqrt{E[S^2] - \{E[S]\}^2}/E[S]$.
If $c_S$, equivalently the variance of message sizes in proportional, increases, $R$ increases.
In this example, the value of $c_S$ is very high because the message size distribution exhibits long-tailed property.

\item {\bf When payload size $\ell_d$ is small}

When payload size decreases, the number of segmented packets per message increases,
resulting in the smaller value of $c_S$. 
Hence, the mean response time $R$ may be smaller if payload size is smaller even though the mean batch size increases.

\item {\bf When payload size $\ell_d$ is small enough}

In this case, the segmented packets, of which size is $\ell_d$ is dominant in all created packets.
Therefore, the queueing system can be approximated as an $\rm{M}^{\rm X}/\rm{D}/1$ queueing model.
Although the value of $c_S$ is almost zero,
the mean batch size is too large,
resulting in large mean response time.

\end{itemize}

From Fig.~\ref{fig: log},
we find that 
the mean response time is more convex in payload sizes
if batch arrival rate is larger.

\section{Conclusion}
\label{sec: conclusion}

This paper proposed the $\rm{M}^{\rm X}/\rm{G}/1$ queueing model
to discuss the effect of payload size on mean response time
when message segmentations occur.
We derived the closed form of mean response time,
given payload size, message size distribution and message arrival rate.
From a numerical result,
we have demonstrated that the mean response time is more convex in payload sizes
if message arrival rate is larger.
in a scenario where Web objects are delivered over a physical link.

The remaining issues include the clarification of the relationship among several parameters to express the optimized payload size,
the extension of our model to noisy links,
and development of payload adaptation algorithm to minimize the mean response time.

\section*{Acknowledgment}

This work was supported by JSPS KAKENHI Grant Number JP15K00139. 


\appendix

\begin{figure}
\centering{
\epsfig{file=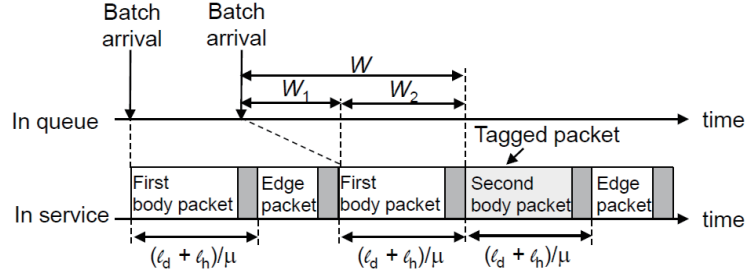, width=10cm} \\
\caption{Relationship between $W_1$ and $W_2$.}
\label{fig: w1w2}
}
\end{figure}

This appendix provides derivations of \eqref{eq: R} -- \eqref{eq: W}.

Equations~\eqref{eq: R} -- \eqref{eq: S2} are clear from the definitions of $R$ and $S$.

To derive \eqref{eq: W},
we observe an arbitrary packet in a batch (i.e., message), referred to as the tagged packet,
and divide the waiting time $W$ into two parts (see Fig.~\ref{fig: w1w2}):
\begin{itemize}

\item waiting time $W_1$: the time from when the first packet in a batch
have the tagged packet arrives at the queueing system
to when it enters service.

\item waiting time $W_2$: the time from when the first packet enters service
to when the tagged packet enters service.

\end{itemize}

From the argument of \cite[Chapter~4]{AKI99},
the form of $E[W_1]$ is given by 
\begin{align}
\label{eq: W1 a}
E[W_1] &=
\cfrac{\lambda \, 
\left(
E[X] \, E\left[S^2\right] 
+
\left(E\left[X^2\right] - E[X] \right) \, \left\{E[S]\right\}^2
\right)
}
{2 \, (1 - a)}.
\end{align}

Noting that one message consists of consecutive $(n - 1)$ body packets and an edge packet
when it it is divided to $n$ packets (see Fig.~\ref{fig: message segmentation}),
we have
\begin{align}
\label{eq: W2 a}
E[W_2] &= \cfrac{\ell_d + \ell_h}{\mu_c} \sum_{n=1}^\infty \, (n-1) \, \Pr. (X = n)
= \cfrac{\ell_d + \ell_h}{\mu_c} 
\left(\sum_{n=1}^\infty \, n \, \Pr. (X = n) n -
\sum_{n=1}^\infty \Pr. (X = n) \right) \notag \\
&= \cfrac{\ell_d + \ell_h}{\mu_c} 
\left(E[X] - 1\right),
\end{align}
because of $\Pr.(X=0) = 0$.
From \eqref{eq: W1 a} and \eqref{eq: W2 a}, we obtain \eqref{eq: W}.
\small

\bibliographystyle{IEEEtran}
\bibliography{paper}

\end{document}